\documentclass[aps,prb,reprint,showpacs,superscriptaddress,longbibliography,floatfix]{revtex4-2}
\usepackage{mathtools,amssymb,graphicx,units}
\usepackage{amsmath,bm}
\usepackage{bbold}

\usepackage[plainpages=false,pdfpagelabels,colorlinks=true,linkcolor=red,urlcolor=blue,citecolor=blue,pdftitle={Title},pdfauthor={},pdfdisplaydoctitle=true,pdfduplex=DuplexFlipLongEdge]{hyperref}
\usepackage{verbatim}
\usepackage[english]{babel}
\usepackage{xcolor}
\usepackage{dsfont}

\usepackage[normalem]{ulem}

\usepackage{braket}

\begin{document}

\title{Phase transitions in the Haldane-Hubbard model}

\author{Wan-Xiu He}
\affiliation{Beijing Computational Science Research Center, Beijing 100084, China}

\author{Rubem Mondaini}
\email{rmondaini@csrc.ac.cn}
\affiliation{Beijing Computational Science Research Center, Beijing 100084, China}

\author{Hong-Gang Luo}
\affiliation{School of Physical Science and Technology, Lanzhou University, Lanzhou 730000, China}
\affiliation{Lanzhou Center for Theoretical Physics $\&$ Key Laboratory of Theoretical Physics of Gansu Province, Lanzhou University, Lanzhou 730000, China}
\affiliation{Beijing Computational Science Research Center, Beijing 100084, China}

\author{Xiaoqun Wang}
\affiliation{School of Physics, Zhejiang University, Hangzhou, 310058, China}

\author{Shijie Hu}
\email[]{shijiehu@csrc.ac.cn}
\affiliation{Beijing Computational Science Research Center, Beijing 100084, China}
\affiliation{Department of Physics, Beijing Normal University, Beijing, 100875, China}

\begin{abstract}
The Haldane-Hubbard model is a prime example of the combined effects of band topology and electronic interaction. We revisit its spinful phase diagram at half-filling as a consensus on the presence of SU($2$) symmetry is currently lacking. To start, we utilize the Hartree-Fock mean-field method, which offers a direct understanding of symmetry breaking through the effective mass term that can acquire spin dependence. Our results, in agreement with previous studies, provide an instructive insight into the regime where the Chern number $C=1$, with only one spin species remaining topological. Besides that, we numerically study the phase diagram of the Haldane-Hubbard model via a large-scale infinite-density matrix renormalization group (iDMRG) method. The phase boundaries are determined by the Chern number and the correlation lengths obtained from the transfer-matrix spectrum. Unlike previous studies, the iDMRG method investigates the Haldane-Hubbard model on a thin and infinitely long cylinder and examines scenarios consistent with the two-dimensional thermodynamic limit. Here, the phase diagram we obtained qualitatively goes beyond the Hartree-Fock scope, particularly in the $C=1$ region, and serves as a quantitative benchmark for further theoretical and experimental investigations.
\end{abstract}
\maketitle

\section{Introduction}\label{sec:intro}

The study of topologically ordered states has garnered significant attention in condensed matter physics, especially since the discovery of the quantum Hall effect (QHE)~\cite{Klitzing1980, Delahaye1986, Klitzing2020}. This field aims to understand the intricate interplay between topology and electronic properties in materials~\cite{Zhang2005, Abanin2011}, leading to the emergence of novel phenomena and potential applications in quantum computing~\cite{Nayak2008} and spintronics~\cite{Smejkal2018, He2022}. Haldane's original proposal suggested that a honeycomb lattice model with electrons hopping on it could exhibit QHE without requiring Landau levels, indicating the potential for non-trivial topology in basic band insulators~\cite{Haldane1988}. This discovery culminated in its experimental realization using ultracold atoms~\cite{Jotzu2014} and sparked a surge of interest in studying systems with both topologically non-trivial band structures and electron-electron interactions. In recent years, there has been a particular focus on the Haldane model~\cite{wang2010charge, Varney2011, Zheng2015, Hickey2015, Wu2016, Imriska2016, Vanhala2016, Mertz2019, Shi2021, Shao2021, Yi2021, Yi2022, Mai2023, Yuan2023}, as well as other models such as the Kane-Mele~\cite{Zheng2011, Hohenadler2012, Budich2012, Bercx2014} and Bernevig-Hughes-Zhang models~\cite{Yoshida2013, Chen2019, Tzeng2023}.

Concerning the spinful Haldane-Hubbard model and its variations, the exploration of the topological phases and phase transitions has been performed by various numerical methods, including the static mean-field (MF) theory~\cite{Zheng2015, Hickey2015, Vanhala2016, Shao2021, Shi2021}, exact diagonalization (ED)~\cite{Vanhala2016, Shao2021, Yuan2023}, dynamical mean-field theory (DMFT)~\cite{Vanhala2016} as well as the cluster variations~\cite{Imriska2016, Wu2016}, and bold diagrammatic Monte Carlo (BDMC) technique~\cite{Tupitsyn2019}. When the ground state exhibits non-trivial topology, a Chern number $C \neq 0$ exists, mapping the number of protected chiral edge modes in the presence of open boundary conditions (OBC). When the interacting (Hubbard) term is turned off, the topological insulator of the spinful Hamiltonian leads to a total Chern number $C=2$ (CI$_2$), wherein each spin species contributes with a Chern number $C=1$. Remarkably, an exotic insulator with a Chern number $C=1$ (CI$_1$) can also emerge, which has been considered as a result of spontaneous spin-rotation symmetry breaking. Although the observation of this state has reached a certain consensus~\cite{Vanhala2016, Tupitsyn2019, Yuan2023}, there are qualitative and quantitative inconsistencies in the phase diagrams obtained by different methods. Moreover, a debate persists regarding the presence of CI$_1$ in the extended Haldane-Hubbard model~\cite{Shao2021, Shi2021}. As such, the phase diagram of the Haldane-Hubbard model and the comprehensive characterization of the associated phase transitions remain largely unresolved.

Here, we mainly focus on the model at half-filling on a thin and infinitely long cylinder to address these questions. The infinite density matrix renormalization group (iDMRG) method is employed to better understand phase transitions in the model. In addition, we revisit the Hartree-Fock method and present results that help set the stage for the unbiased ones from iDMRG and can be used to readily understand its mapping onto the familiar non-interacting phase diagram of the model~\cite{Haldane1988}. Furthermore, we investigate the symmetry-broken topological phase CI$_1$ with a Chern number $C=1$.

Our paper is structured as follows. In Sec.~\ref{MODEL AND METHOD}, we introduce the Haldane-Hubbard model on a honeycomb lattice and provide a detailed explanation of the two distinct lattice structures that were examined using the iDMRG simulations. In Sec.~\ref{Mean-field analysis}, we present the Hartree-Fock approximation and results, which are used to benchmark and introduce all phases and phase transitions. Section~\ref{iDMRG calculations} showcases our main results, specifically the phase diagram at half-filling with the iDMRG method. Lastly, we conclude our paper in Sec.~\ref{CONCLUSIONS}. The appendices contain additional numerical results for other lattice structures obtained through the iDMRG method, as well as other parameter sets in mean-field results.

\section{MODEL AND METHOD}\label{MODEL AND METHOD}

\begin{figure}[!htbp]
\centering
\includegraphics[width=0.5\textwidth]{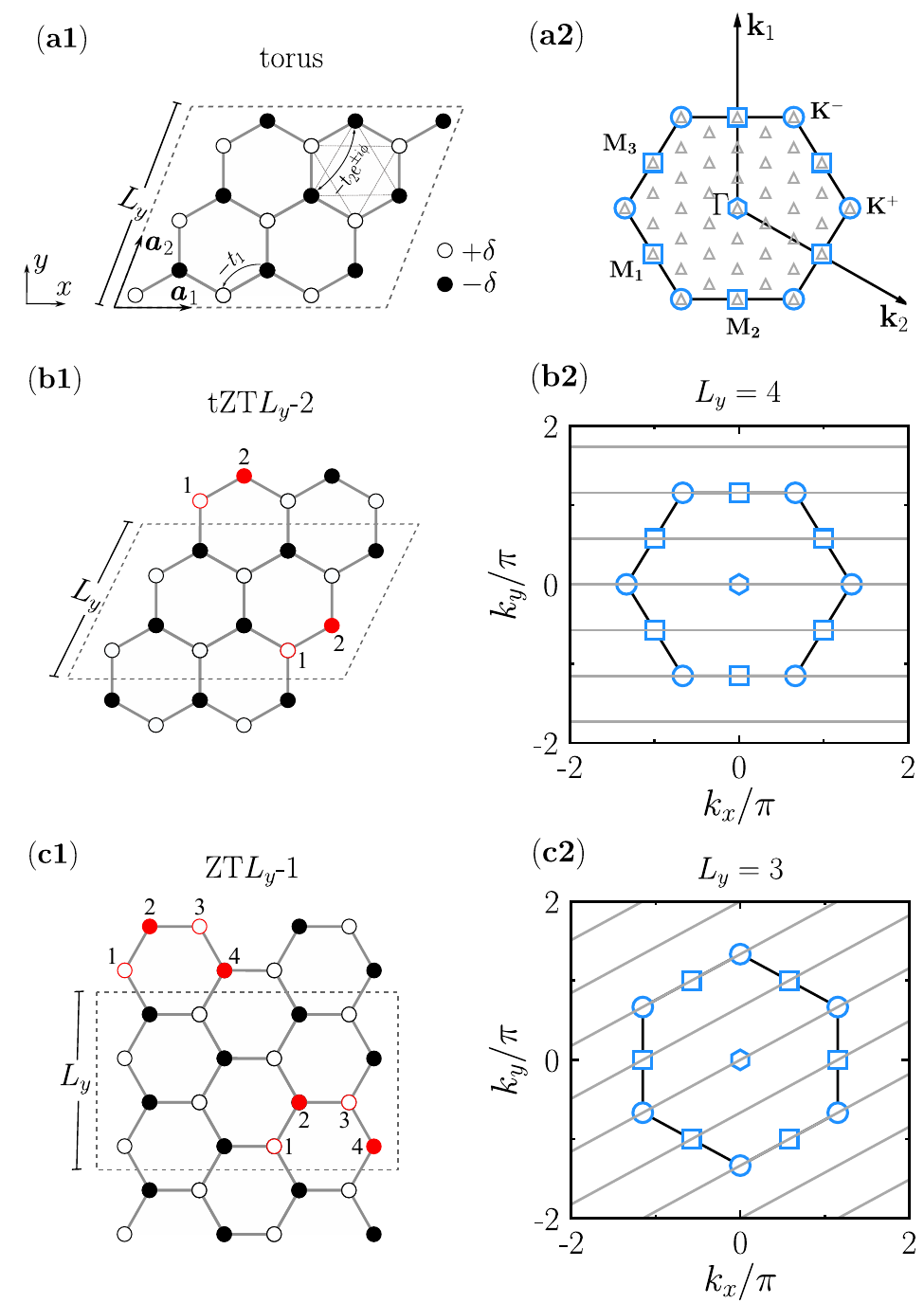}
\caption{(a1) Schematic cartoon marking the relevant terms in the Hamiltonian, where the NNN links get a phase $e^{+{\rm i}\phi}$ ($e^{-{\rm i}\phi}$) for clockwise (counterclockwise) directions. This lattice shape is used in the mean-field calculations, here shown for a linear size $L_x = L_y = L = 3$. (a2) FBZ for the diamond-shaped cluster and corresponding valid momenta in the case of $L=6$. The lattice structures for two distinct cylinders (b1) tZT$L_y$-$2$ and (c1) ZT$L_y$-$1$ used in the iDMRG simulations. Both structures have a rolling $n$-site shift, as marked by the red lattice sites. (b2) According to the condition \eqref{eq:momentum_selection}, valid momentum cutting lines (grey), that thread FBZ, are plotted for the cylinder tZT$4$-$2$. Similarly, (c2) for ZT$3$-$1$. The high-symmetry points, $\bf \Gamma$, ${\bf M}_1$, ${\bf M}_2$, ${\bf M}_3$, and ${\bf K}^\pm$, in (a2-c2) are annotated.}
\label{fig:fig_1}
\end{figure}

On a honeycomb lattice composed of two nested triangular sublattices, A and B, the Hamiltonian of the Haldane-Hubbard model (HHM) reads
\begin{eqnarray}
\hat{H} &&=-t_1 \sum_{\braket{l,l'},\sigma} \hat{c}_{l\sigma}^{\dag} \hat{c}_{l'\sigma}^{\phantom{\dag}} - t_2\sum_{\langle\!\langle l,l'\rangle\!\rangle ,\sigma} e^{{\rm i}\phi_{ll'}} \hat c_{l\sigma}^{\dag} {\hat c}_{l'\sigma}^{\phantom{\dag}}\nonumber\\
&&+U\sum_{l}\left({\hat n}^{\phantom{\dag}}_{l\uparrow}-\frac{1}{2}\right)\left({\hat n}^{\phantom{\dag}}_{l\downarrow}-\frac{1}{2}%
\right)+\delta\sum_{l,\sigma} s^{\phantom{\dag}}_l \hat{n}^{\phantom{\dag}}_{l\sigma},
\label{eq:model}
\end{eqnarray}
where $\hat{c}_{l\sigma}^{\dag}$ ($\hat c_{l\sigma}^{\phantom{\dag}}$) is the creation (annihilation) operator for an electron with spin $\sigma=\uparrow$, $\downarrow$ at site-$l$, $\hat{n}^{\phantom{\dag}}_{l\sigma}=\hat{c}^{\dag}_{l\sigma} {\hat c}^{\phantom{\dag}}_{l\sigma}$ the particle number operator, $t_1$ and $t_2$ the hopping amplitudes on the nearest-neighbor (NN) and next-nearest-neighbor (NNN) links respectively, $U > 0$ the strength of the onsite repulsion, and $\delta > 0$ the staggered chemical potential with the sign $s^{\phantom{\dag}}_l = +1$ ($-1$) for the sublattice A (B). Lastly, $l$ sums over $N$ unit cells or equivalently $2N$ sites; $\braket{l,l'}$ and $\langle\!\langle l, l'\rangle\!\rangle$ run over all NN and NNN links -- see Fig.~\ref{fig:fig_1}(a)

In this model, the time-reversal symmetry is broken due to the presence of complex phase factors $e^{{\rm i}\phi_{ll'}}$. Specifically, the sign of the phase angle $\phi_{ll'}$ depends on the hopping direction on the triangular lattice consisting of NNN links, with a positive (negative) value for the clockwise (counterclockwise) hopping. Hereafter, $t_1=t$ sets the energy unit, and we choose $t_2 = 0.2t$, and $|\phi_{ll'}|=\phi=\pi/2$. Additionally, we focus our calculations on the half-filling, meaning there is one electron per site on average. This is represented by $N_{\uparrow} = N_{\downarrow} = N$, where $N_{\sigma} = \sum_{l} \braket{\hat{n}^{\phantom{\dag}}_{l\sigma}}$ denotes the number of electrons for species-$\sigma$.

To start, we describe a few symmetries of the model. First, the Hamiltonian (\ref{eq:model}) is invariant under an arbitrary choice of phase $\varphi_{\sigma}$, i.e.,
\begin{eqnarray}
\hat{c}^{\phantom{\dag}}_{l\sigma} \rightarrow \hat{c}^{\phantom{\dag}}_{l\sigma} e^{{\rm i} \varphi_{\sigma}} \quad \text{and} \quad \hat{c}^{\dag}_{l\sigma} \rightarrow {\hat c}^{\dag}_{l\sigma} e^{-{\rm i}\varphi_{\sigma}}\ ,
\end{eqnarray}
denoting a conservation of U(1) symmetry, i.e., the number of electrons $N_{\sigma}$ for each spin species. Besides that, under a rotation
\begin{equation}
\mathcal{R}(\varphi_r)=
\begin{pmatrix}
\phantom{-}\cos{\varphi_r} & \sin{\varphi_r}\\
-\sin{\varphi_r} & \cos{\varphi_r}
\end{pmatrix}
\end{equation}
with an arbitrary angle $\varphi_r$, we get a new pair of annihilation operators $(\hat{c}_{l\uparrow}^\prime, \hat{c}_{l\downarrow}^\prime)^\text{T} = \mathcal{R}(\varphi_r) (\hat{c}^{\phantom{\dag}}_{l\uparrow}, \hat{c}^{\phantom{\dag}}_{l\downarrow})^\text{T}$, while the Hamiltonian remains unchanged. Combining the above-mentioned double U($1$) symmetries with the rotation symmetry yields a total symmetry of the model \eqref{eq:model}, characterized by the U($1$)$\times$SU($2$) group for the charge and spin degrees of freedom, separately. In particular, when $\varphi_r = \pi/2$, the rotation $\mathcal{R}(\pi/2)$ gives
\begin{eqnarray}
\hat{c}^\prime_{l\uparrow} = \hat{c}^{\phantom{\dag}}_{l\downarrow} \quad \text{and} \quad \hat{c}^\prime_{l\downarrow} = -\hat{c}^{\phantom{\dag}}_{l\uparrow}\ .
\end{eqnarray}
This corresponds to the ``time-reversal" operation defined in the space of spin degrees of freedom, which is anti-unitary, i.e., $\mathcal{R}^2(\pi/2) = -\mathbb{1}$, so we call it the {\it spin time-reversal} symmetry. Lastly, particle-hole symmetry is also preserved at half-filling. Regardless of whether the staggered potential and the phase $\phi$ change sign, the overall physics remains unchanged~\cite{Vanhala2016}.

On the honeycomb lattice, each sublattice site A has three neighboring sublattice sites B, connected by vectors ${\bf c}_{1}= (0$, $\sqrt{3} / 3)$, ${\bf c}_{2}=(1 / 2$, $-\sqrt{3} / 6)$, ${\bf c}_{3}=(-1 / 2$, $-\sqrt{3} / 6)$ in two-dimensional coordinates, where $\vert {\bf c}_{1} \vert = \vert {\bf c}_{2} \vert = \vert {\bf c}_{3} \vert = \sqrt{3}/3$ -- see Fig.~\ref{fig:fig_1}(a1). We can roll the honeycomb lattice onto a sphere if three edges along ${\bf c}_{1}$, ${\bf c}_{2}$ and ${\bf c}_{3}$ directions share an equal side length. After choosing primitive vectors
\begin{eqnarray}
\bm{a}_{1} = \left( \frac{1}{2},\ \frac{\sqrt{3}}{2} \right) \quad \text{and} \quad \bm{a}_{2} = \left( 1,\ 0 \right)\ ,
\end{eqnarray}
the first Brillouin zone (FBZ), expanded in reciprocal vectors
\begin{eqnarray}
\bm{b}_{1} = \left( 0,\ \frac{4\pi}{\sqrt{3}} \right) \quad \text{and} \quad \bm{b}_{2} = \left( 2\pi,\ -\frac{2\pi}{\sqrt{3}} \right)\ ,
\end{eqnarray}
has high-symmetry points relevant to the point group symmetry $D_{6h}$ of the lattice: two inequivalent $K$-points ${\bf K}^{\pm}$, one $\Gamma$-point ${\bf \Gamma} = (0, 0)$, three distinct $M$-points (i.e., ${\bf M}_1$, ${\bf M}_2$ and ${\bf M}_3$) - see Fig.~\ref{fig:fig_1}(a2). As both $\delta$ and $\phi$ are finite in the Hamiltonian \eqref{eq:model}, the $D_{6h}$ symmetry reduces to a lower $C_{3}$ one, specifically $120^\circ$ rotational symmetry. However, on a cylinder that we study by the iDMRG method~\cite{PhysRevLett.69.2863, RevModPhys.77.259, PhysRevB.48.10345, mcculloch2008infinite, PhysRevB.87.235106}, the $C_{3}$ symmetry is also absent.

To carry out the iDMRG simulations, one obtains the cylindrical geometry by rolling the sheared two-dimensional lattice onto a thin and infinitely long cylinder. The allowed cutting lines in the FBZ encompass the momentum points where the gap closure of the low-energy excitations is most relevant for capturing phase transitions. This procedure is specifically performed when high-symmetry points are available~\cite{Varney2011, Shao2021, Yuan2023}. Figures~\ref{fig:fig_1}(b1, c1) illustrate this process for two different cylinders: zigzag-top (ZT) and tilted-zigzag-top (tZT), respectively, distinguished by the orientation along the short direction and the way they are rolled up. For example, we plot the cylinder tZT$L_y$-$2$ in Fig.~\ref{fig:fig_1}(b1), which has zigzag short-edges and a shift of two sites. For the cylinder tZT$L_y$-$n$, a translation by $L_{y}$ sites along the short circumference ($y$-axis) is equivalent to a translation by $n$ sites along the $x$-axis, which selects the momentum ${\bf k}=(k_x$, $k_y)$ as follows:
\begin{eqnarray}
L_y \left( \frac{k_x}{2} + \frac{\sqrt{3} k_y}{2} \right) = 2 \pi p + n k_x\ ,\label{momentumline}
\label{eq:momentum_selection}
\end{eqnarray}
with an arbitrary integer $p$.  Each selected value of $p$ gives a cutting line of momentum in the FBZ. Previous studies~\cite{Shao2021, PhysRevB.102.121102} have shown that various cylinders with distinct $L_{y}$ and $n$ yield momentum cutting lines in FBZ. This work mainly considers two cylinders, that is, ZT$3$-$1$ and tZT$4$-$2$. As illustrated in Fig.~\ref{fig:fig_1}(b2, c2), these strike a balance of encompassing all the necessary high-symmetry points relevant to the phase transitions, i.e., two $K$-points, one $\Gamma$-point, and three $M$-points, while having sufficiently small variational unit cell (encoded in the values  $n=1$ and $2$) that result in amenable computing times in iDMRG. In Appendix~\ref{app:TZT2-1}, we also study one extra geometry, the cylinder tZT$2$-$1$, which we show to be significantly impacted by finite-size effects precisely because some of the high-symmetry points are unavailable. In Sec.~\ref{correlationlengthspectrum}, we further analyze the momentum resolution of the correlation-length spectra for cylinders tZT$L_y$-$1$. 

\section{Mean-field analysis}
\label{Mean-field analysis}
\begin{figure}[!htb]
\centering
\includegraphics[width=1\linewidth]{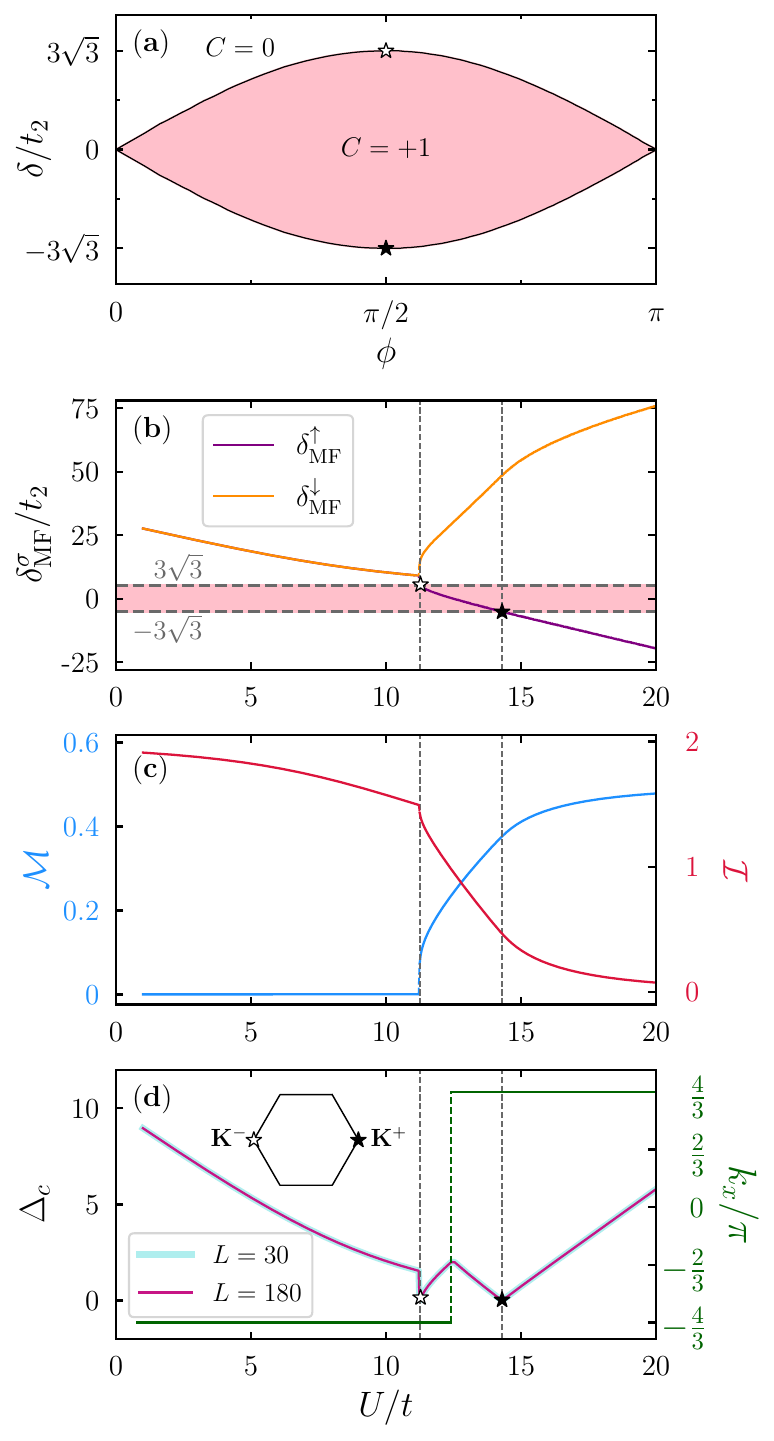}
\caption{Mean-field results for staggered potential $\delta / t = 6$. (a) The positive-$\phi$ phase diagram in the non-interacting limit ($U=0$) for each spin species. (b) The effective staggered potential $\delta^\sigma_{\rm MF}$ for $\sigma = \uparrow$, $\downarrow$ defined in \eqref{effCPMF}, as a function of the interaction strength $U$. The shaded region marks the topological regime. (c) The staggered magnetization order parameter $\mathcal{M}$ and the sublattice density imbalance $\cal I$ as a function of $U / t$. (d) The charge gap $\Delta_c ({\bf k})$ defined in \eqref{eq::chargegap} as a function of $U/t$. Here, contrasting two different system sizes, $L=30$ and $180$ -- finite-size effects are marginal. Inset: FBZ with two $K$-points $\mathbf{K}^\pm$ annotated [and similarly shown in (a) and (b)], and where a gap closing occurs at values of the interactions $U_{c1} / t = 11.28$ and $U_{c2} / t = 14.3$, respectively, marking the two topological transitions. The right vertical axis gives the location $k_x$ of the smallest gap $\Delta_c$.}
\label{fig:fig_2}
\end{figure}

Let us begin by describing some general features of the original Hamiltonian~\eqref{eq:model}. Setting $U=0$, the model reduces to the original Haldane model. In the case of $\delta=0$, the phase $\phi=\pi/2$ guarantees a Chern insulator CI$_2$ with a Chern number $C=2$ in the ground state ($C_\sigma = 1$ for each spin species). Such Chern insulator survives until the closure of the gap at $|\delta/t_2| = 3\sqrt{3}$, evolving into a band-insulator with a trivial charge density wave (CDW) state, manifesting a finite density imbalance between sublattices A and B~\cite{Haldane1988}, given by
\begin{eqnarray}
{\cal I} = \left\lvert n_{\text{A}\uparrow} + n_{\text{A}\downarrow} - n_{\text{B}\uparrow} - n_{\text{B}\downarrow} \right\rvert\ ,
\end{eqnarray}
where
\begin{eqnarray}
\left\{\begin{array}{cl}
n_{\text{A}\uparrow} = \braket{\hat{n}_{l\uparrow}} \ \text{and} \ n_{A\downarrow} = \braket{\hat{n}_{l\downarrow}} & \quad \text{for} \quad l \in \text{A}\\\\
n_{\text{B}\uparrow} = \braket{\hat{n}_{l\uparrow}} \ \text{and} \ n_{B\downarrow} = \braket{\hat{n}_{l\downarrow}} & \quad \text{for} \quad l \in \text{B}
\end{array}
\right.\ ,
\end{eqnarray}
denote the number of electrons for distinct sublattice sites and species. We stress that ${\cal I} \ne 0$ once $\delta > 0$ and thus the transition at $\lvert \delta / t_{2} \rvert = 3\sqrt{3}$ occurs without any relevant symmetry breaking.

In the other limit of $\delta=0$ and $U\gg t_{1}$, $t_{2}$, the presence of charge excitations is suppressed: The second-order hopping processes result in an effective Heisenberg model, which includes NN and NNN antiferromagnetic Heisenberg couplings, denoted as $J_1$ and $J_2$, respectively. When the ratio of $J_{2} / J_{1} = t^{2}_{2} / t^{2}_{1} \lesssim 0.27$~\cite{Mezzacapo2012, Gong2013, Gong2015}, a N\'{e}el phase can be achieved in the honeycomb lattice. In this phase, the ground state exhibits SU($2$) symmetry breaking, and the low-energy excitation spectrum contains gapless Goldstone modes~\cite{Goldstone1962, Low2002}. Nonetheless, when hopping energy scales are compatible with the repulsive interactions, the emergence of CI$_1$ with spontaneous spin-rotation symmetry breaking has also been observed in this model using various methods~\cite{Vanhala2016, Tupitsyn2019, Yuan2023}.

In the simplest mean-field treatment, the on-site interaction is approximated by
\begin{eqnarray}
{\hat n}_{l\uparrow}{\hat n}_{l\downarrow} \approx {\hat n}_{l\uparrow} \braket{\hat{n}_{l\downarrow}} + \braket{\hat{n}_{l\uparrow}}
{\hat n}_{l\downarrow} - \braket{\hat{n}_{l\uparrow}} \braket{\hat{n}_{l\downarrow}}\ .
\end{eqnarray}
This approximation assumes that one species of electrons move in the background of the site-dependent chemical potential provided by the other species. The superconducting term $\braket{\hat{c}_{l\uparrow} \hat{c}_{l\downarrow}}$, which breaks U($1$) symmetry, is also neglected here because $U>0$, following the BCS paradigm~\cite{PhysRevB.36.857, PhysRevB.62.9083}. We notice that spin-flipping terms $\braket{\hat{c}^{\dagger}_{l\uparrow} \hat{c}^{\phantom{\dagger}}_{l\downarrow}}$, describing gapless transverse spin excitations in the N\'{e}el phase, would artificially assign a finite mass to the relevant Goldstone modes, indirectly breaking SU($2$) symmetry. Therefore, the N\'{e}el phase is characterized by the spin-spin correlation function along the $z$-axis, given by
\begin{eqnarray}
S_{\text{SDW}} = \frac{1}{N^2} \sum_{l,l^{\prime}} \braket{ (\hat{n}_{l\uparrow}-\hat{n}_{l\downarrow})  (  \hat{n}_{l^{\prime}\uparrow}-\hat{n}_{l^{\prime}\downarrow}) }\ .
\end{eqnarray}
This correlation function describes the long-range spin staggering in space, corresponding to a spin density wave (SDW) state. However, we observe that including spin-flipping terms in the mean-field approximation does not alter our conclusions below, and we remove them for simplicity.

In principle, a total of $2N$ variational parameters, specifically $\langle \hat{n}_{l\uparrow} \rangle$ and $\langle \hat{n}_{l\downarrow} \rangle$ for all sites, are necessary. In practice, considering the translation symmetry, only four variational parameters, that is, $n_{\text{A}\uparrow}$, $n_{\text{A}\downarrow}$, $n_{\text{B}\uparrow}$, and $n_{\text{B}\downarrow}$, are sufficient for representing the density imbalance of charge and spin between the two inequivalent sublattices in four distinct phases of the quantum phase diagram~\cite{Vanhala2016}.

In reciprocal space, the creation operators are defined as follows
\begin{eqnarray}
\left\{
\begin{array}{cc}
\hat{c}^\dag_{l\sigma} = \frac{1}{\sqrt{N}} \sum_{{\bf k} \in \text{FBZ}}\hat{d}^\dag_{{\bf k} \sigma} e^{-{\rm i}{\bf k}\cdot\mathbf{r}_l} & \quad \text{for} \quad l \in \text{A}\\\\
\hat{c}^\dag_{l\sigma} = \frac{1}{\sqrt{N}} \sum_{{\bf k} \in \text{FBZ}}\hat{g}^\dag_{{\bf k} \sigma} e^{-{\rm i}{\bf k}\cdot\mathbf{r}_l} & \quad \text{for} \quad l \in \text{B}
\end{array}
\right.
\end{eqnarray}
with $\hat{d}^\dag_{\mathbf{k}\sigma}$ and $\hat{g}^\dag_{\mathbf{k}\sigma}$ denoting the creation operators in reciprocal space. The resulting mean-field Hamiltonian is given by
\begin{eqnarray}
\hat{H}^\text{MF} = \sum_{{\bf k} \in \text{FBZ}} \hat{\psi}_{{\bf k}}^{\dag} H^\text{MF}_{\bf k} \hat{\psi}_{\bf k}\ ,
\end{eqnarray}
where $\hat{\psi}_{\bf k}^\dag = ( \hat{d}^\dag_{{\bf k}\uparrow}$, $\hat{g}^\dag_{{\bf k}\uparrow}$, $\hat{d}^\dag_{{\bf k}\downarrow}$, $\hat{g}^{\dag}_{{\bf k}\downarrow} )$ is the spinor notation used as a basis for each lattice momentum ${\bf k}$. The $4 \times 4$ matrix $H^\text{MF}_{\bf k}$ is given by
\begin{equation}
\left(
\begin{array}[c]{cccc}
m^{+}_{{\bf k}} +\delta_{\text{A}\downarrow} & f_{{\bf k}} & & \\
f^{\ast}_{{\bf k}}   & m^{-}_{{\bf k}} + \delta_{\text{B}\downarrow} & & \\
 & & m^{+}_{{\bf k}} +\delta_{\text{A}\uparrow} & f_{{\bf k}} \\
 & & f^{\ast}_{{\bf k}} & m^{-}_{{\bf k}} + \delta_
{\text{B}\uparrow}
\end{array}
\right)
\end{equation}
with the NN structure factor $f_{{\bf k}} = t_{1} (1 + e^{-i{\bf k}\cdot\mathbf{a}_{1}} + e^{-i{\bf k}\cdot\mathbf{a}_{2}})$, the NNN ones $m^{\pm}_{{\bf k}} = \pm 2t_{2} (\sin({\bf k}\cdot\mathbf{a}_{1})-\sin({\bf k}\cdot\mathbf{a}_{2})-\sin({\bf k} \cdot (\mathbf{a}_{1}-\mathbf{a}_{2})))$,
and finally, $\delta_{\text{A} \sigma} = \delta + U n_{\text{A} \bar{\sigma}}$ and $\delta_{\text{B}\sigma} = -\delta + U n_{\text{B} \bar{\sigma}}$. This leads to an effective spin-dependent staggered potential
\begin{eqnarray}
\delta^\sigma_\text{MF} \equiv \frac{1}{2} \left( \delta_{\text{A} \sigma} - \delta_{\text{B} \sigma} \right) = \delta + \frac{U}{2} \left( n_{\text{A} \bar{\sigma}} - n_{\text{B} \bar{\sigma}} \right)\ ,\label{effCPMF}
\end{eqnarray}
where $\bar{\sigma}$ denotes the reverse of $\sigma$.

In the nonlinear variational process, to avoid the problem of metastable states, we minimize the energy by starting from various initial guesses of the ground state. More concretely, each of the four parameters in an initial guess can be randomly chosen from the range of $[0,\ 1]$. However, given that the exchanging symmetry between species potentially breaks, we restrict the initial states to have $n_{\text{A}\uparrow} > n_{\text{A}\downarrow}$ and $n_{\text{B}\downarrow} > n_{\text{B}\uparrow}$~\footnote{The rationale behind this choice is that if a constraint was not imposed, the behavior of the two spin-species are often reversed in what concern their topological characteristics and a smooth variation of $\delta_\text{MF}^\sigma$ is not seen, but rather a rapid oscillation when increasing $U/t$ between the current values shown in Fig.~\ref{fig:fig_2}(b)}. Apart from the normal CDW state, the staggered magnetization is represented by $\mathcal{M} = \sqrt{ \mathcal{M}^2_x + \mathcal{M}^2_y + \mathcal{M}^2_z }$, where
\begin{eqnarray}
\mathcal{M}_{x/y/z} = \frac{1}{2} \sqrt{\left\langle \left(\hat{S}^{x/y/z}_\text{A} - \hat{S}^{x/y/z}_\text{B}\right)^2 \right\rangle}
\end{eqnarray}
represent the components of staggered magnetization in the x, y, and z-axes,
\begin{eqnarray}
\hat{\mathbf{S}}_{\text{A}/\text{B}} = \sum_{\sigma \sigma^\prime} \hat{c}_{l \sigma}^\dag \mathbb{S}^{\phantom{\dag}}_{\sigma\sigma^\prime} \hat{c}^{\phantom{\dag}}_{l \sigma^\prime}
\end{eqnarray}
gives the vector operators of the spin at site-$l$ belonging to the sublattices A and B, and the vector $\mathbb{S} = (S^x$, $S^y$, $S^z)$ includes the matrix representations $S^x$, $S^y$ and $S^z$ for the spin-$1/2$ operators.

To characterize the different phases and associated transitions, we now consider the case of fixed $\delta / t = 6$ in Fig.~\ref{fig:fig_2} as an example. At small interactions $U$, due to the strong alternating chemical potential, electrons mostly occupy sublattice B, resulting in ${\cal I} > 0$ [Fig.~\ref{fig:fig_2}(c)], namely $n_{\text{B}\sigma} \gg n_{\text{A}\sigma}$. Besides that, no staggered spin polarization is observed, resulting in $\mathcal{M} = 0$; such a regime corresponds to a normal CDW state with a band gap. In contrast, the ground state favors a Mott insulator with SDW in the large-$U$ region, where double occupancies are disfavored and spin-up/down electrons separately reside on sublattices A/B. This results in a reduced imbalance ${\cal I}\to 0$ while $\mathcal{M}$ turns finite.

Interestingly, a topologically nontrivial Chern insulator with finite staggered spin polarization emerges in an intermediate region, the spin-rotation symmetry-broken CI$_1$ state. This is easy to see by mapping to the effective staggered potential for each spin species $\delta_\text{MF}^{\sigma}$.
Compared to the phase diagram of the original non-interacting Haldane model~\cite{Haldane1988}, because of the effective chemical potential [Fig.~\ref{fig:fig_2}(b)]
\begin{eqnarray}
\left\lvert \delta_\text{MF}^\uparrow / t_2 \right\rvert < 3 \sqrt{3} \quad \text{and} \quad \delta_\text{MF}^\downarrow / t_2 > 3\sqrt{3}\ ,
\end{eqnarray}
only spin-up species contributes with a finite Chern number $C=1$ [see Fig.~\ref{fig:fig_2}(a)].

Two consecutive topological transitions, CDW-CI$_1$  and CI$_1$-SDW, occur as the interaction strength $U$ increases. These transitions can be characterized by the closure of the charge gap
\begin{eqnarray}
\Delta_c ({\bf k}) = \min_{{\bf k}_1, {\bf k}_2}[E_2({\bf k}_2) - E_1({\bf k}_1)]\ ,\label{eq::chargegap}
\end{eqnarray}
which is defined as the minimum energy difference between the first unoccupied band $E_2$ at momentum ${\bf k}_2$ and the last occupied band $E_1$ at momentum ${\bf k}_1$. They occur approximately at transition points $U_{c1} / t\simeq11.28$ and $U_{c2}/t\simeq14.3$, with gap closing at two distinct Dirac points ${\bf k} = {\bf k}_2 - {\bf k}_1 = {\bf K}^\pm$, respectively. The magnetic order parameter $M$ behavior suggests a first-order transition for CDW-CI$_1$, while the CI$_1$-SDW transition appears to be continuous. We consider a fully periodic diamond-shaped lattice with a linear dimension $L$ to study these transitions. When $L$ is a multiple of 3, all the high-symmetry momentum points are captured, as illustrated in Fig.~\ref{fig:fig_1}(a2) for $L=6$.

In summary, we obtain the phase diagram within the mean-field approximation in the parameter space $(U,\ \delta)$ quantitatively agreeing with existing results~\cite{Vanhala2016}. Here, at large $\delta$, there are three distinct phases with increasing the interaction strength $U$: band-insulating with trivial CDW in the weak-$U$ region, Chern insulator CI$_1$ with SDW in the intermediate region, trivial SDW in the strong-$U$ region. If the original staggered potential is small, one also recovers a Chern insulator CI$_2$ phase when the interaction strength $U$ is sufficiently small (see Appendix~\ref{app:mf_app}).
In the next section, we depart from this approximation, considering all quantum fluctuations, and advance that the microscopic physical scenario of the spin part qualitatively changes in the CI$_1$ and SDW regions.

\section{iDMRG calculations}\label{iDMRG calculations}
Having established the mean-field results, particularly on the emergence of the symmetry-broken phase with a Chern number $C=1$, in this section, we use the iDMRG method to supplement them. In what follows,  Secs.~\ref{Correlation lengths} and \ref{Chern number}  are used to briefly introduce the algorithm for evaluating characteristic correlation lengths and the Chern number, respectively, which are then employed to identify the different phases of the Hamiltonian \eqref{eq:model} in Sec.~\ref{Phase transitions}.

\subsection{Correlation lengths}\label{Correlation lengths}
In the iDMRG method for an infinitely long cylinder, a ``snake-like" matrix product state (MPS) is translation invariant along the longitudinal direction~\cite{He2017, Hu2019}. Within this representation, we can immediately obtain the transfer matrix for a large variational unit cell along the circumference, whose eigenvalues indicate the contribution of different excitations to the correlation functions~\cite{SCHOLLWOCK201196, Zauner_2015}. By normalizing the MPS, we obtain the dominant eigenvalue $\gamma_\text{max}=1$ and sort the eigenvalues in descending order of their amplitudes.

Using U($1$) symmetry for each spin species individually, we divide the transfer matrix into multiple subspaces. For a subspace $\cal S$, we get conserved numbers of electrons for each spin species $N^{\cal S}_\uparrow$ and $N^{\cal S}_\downarrow$. By taking the ground state at half-filling as a reference, we can define the deviation in the number of electrons and the spin polarization along the z-axis as follows
\begin{eqnarray}
\delta N = N^{\cal S}_{\uparrow} + N^{\cal S}_{\downarrow} - 2 N \ \text{and} \ S^z = \frac{1}{2} \left( N^{\cal S}_{\uparrow} - N^{\cal S}_{\downarrow} \right)\ . \quad
\end{eqnarray}

Within the subspace $\cal S$ of the transfer matrix, which is labeled by $(\delta N, S^z)$, the $j$-th eigenvalue can be expressed as
\begin{eqnarray}
\gamma^{\cal S}_j = \exp \left({\rm i} k^{\cal S}_{j} - \left \lvert \mathbf{a}_x \right\rvert / \xi^{\cal S}_{j} \right)\ ,\label{tmeigval}
\end{eqnarray}
where real numbers $k^{\cal S}_{j}$ and $\xi^{\cal S}_{j}$ represent the momentum and correlation length, respectively, and $\mathbf{a}_x$ is the translation vector between two NN variational unit cells.
For cylinders types tZT$L_y$-$n$ and ZT$L_y$-$n$ described in Sec.~\ref{MODEL AND METHOD}, $\left\vert \mathbf{a}_x \right\vert = 1$ and $\sqrt{3}$, respectively. Thus, the correlation lengths in the subspace $\cal S$ can be defined as
\begin{eqnarray}
\xi^{\cal S}_{j} = -\vert\mathbf{a}_x\vert/\ln\left\vert\gamma^{\cal S}_{j}\right\vert\ .
\end{eqnarray}
For the case of $n=1$, the variational unit cell in practice reduces to one $L_y$-th of the above-mentioned size, i.e., $2$ and $4$ for the geometry tZT$L_y$-$1$ and ZT$L_y$-$1$, respectively. In such cases, we have
\begin{eqnarray}
\xi^{\cal S}_{j} = -\vert\mathbf{a}_x\vert / (L_y\ln\left\vert\gamma^{\cal S}_{j}\right\vert)\ .
\end{eqnarray}

Here, we focus on three representative correlation lengths in the subspaces specified by ${\cal S} = (0, 0)$, $(1, 1/2)$, and $(0, 1)$:
$\xi_c \equiv \xi^{(1, 1/2)}_0$, associated with the charge excitation where a single spin-up electron is added; $\xi_s \equiv \xi^{(0, 1)}_0$, associated with the spin-flipping excitation where a spin-up electron is added after a spin-down electron is removed; and $\xi_n \equiv \xi^{(0, 0)}_1$, associated with a neutral excitation that potentially indicates a combination of the charge and spin excitations.

In the ground-state phase diagram of HHM, the lowest-energy charge excitations in all phases have a finite gap, implying the insulating nature of all phases,  topologically trivial or not. Assuming that they are Lorentz-invariant, the inverse of the correlation length $1 / \xi_c$ is proportional to the charge gap $\Delta_c$. At a continuous phase transition point, where the charge gap closes in the thermodynamic limit, the relevant correlation length $\xi_c$ becomes infinite. However, due to the finite truncated bond dimension $m$ and the small circumference of the cylinder, sharp peaks in $\xi_c$ arise instead.
Besides that, the spin and neutral excitations have an ``extra degeneracy'' in the CI$_1$ state, specifically a spin-triplet state with degenerate $S^z = 0$, $\pm1$ states, where $\xi_s = \xi_n$, exemplified with cylinder tZT$2$-$1$ in Appendix~\ref{app:TZT2-1}.
That degeneracy implies that the lowest-energy elementary excitation behaves as a magnon carrying a total spin of $1$, similar to the Goldstone modes in the SDW phase (or precisely, N\'{e}el phase)~\cite{Albuquerque_2011}. Therefore, when $U$ is fixed, we determine the phase boundaries according to two features in the correlation lengths as a function of $\delta$: the sharp peaks of $\xi_c$ and the overlaps between $\xi_s$ and $\xi_n$.

\begin{figure*}[t!]
\centering
\includegraphics[width=0.99\textwidth]{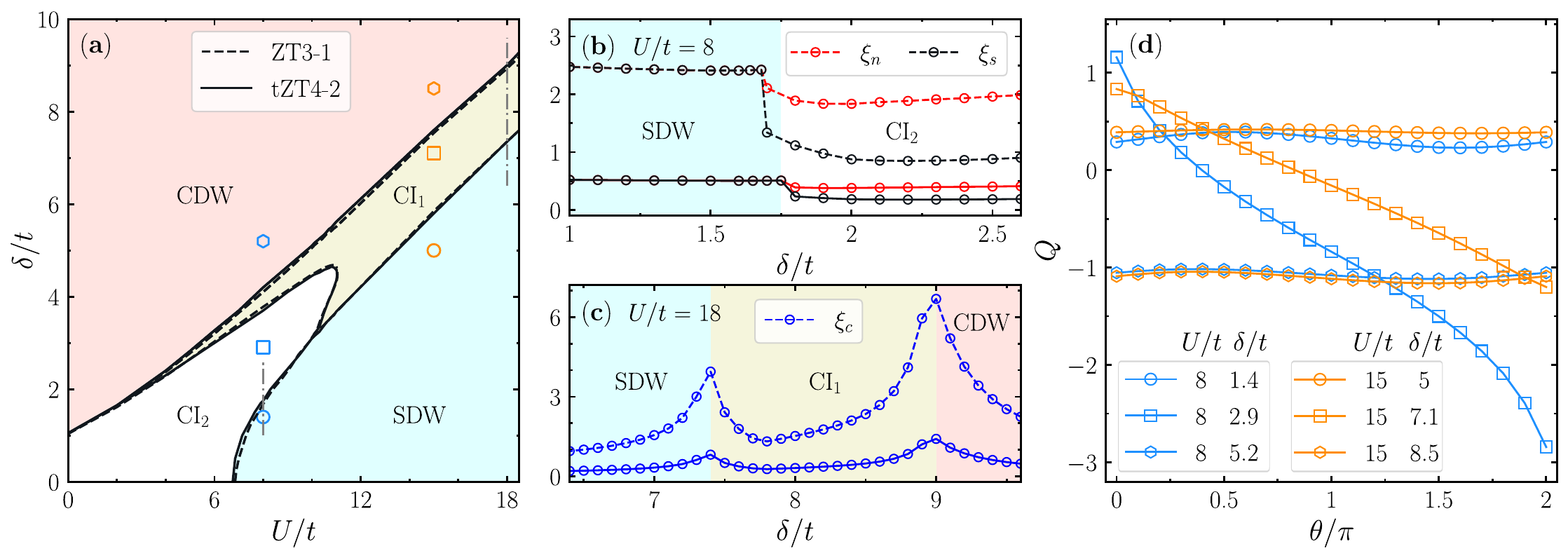}
\caption{(a) Phase diagram of the HHM \eqref{eq:model} in the $(U,\delta)$ space of parameters determined by the iDMRG simulations with a cylindrical geometry and a truncated bond dimension $m=2048$. Here we contrast the phase boundaries calculated from two different cylinders ZT$3$-$1$, and tZT$4$-$2$ exhibiting overall minimal discrepancy. (b) Correlation lengths for neutral ($\xi_n$) and spin ($\xi_s$) excitations for $U/t=8$, showing the SDW-CI$_2$ transition. (c) The charge excitation correlation length ($\xi_c$) at $U/t=18$ whose peaks mark the SDW-CI$_1$ and CI$_1$-CDW transitions. In (a), (b), and (c), dashed (solid) lines are data for the ZT$3$-$1$ (tZT$4$-$2$) cylinder with $m=4096$. The colored regions in (b) and (c) represent the corresponding phases classified for the tZT$4$-$2$ cylinder in (a). (d) Accumulative discrepancy $Q$ as a function of the adiabatically inserted magnetic flux $\theta$ with a small interval $\delta\theta/\pi=0.1$ for the tZT$4$-$2$ cylinder, using a truncated bond dimension $m = 4096$. The markers and color codes are also inserted in (a).}
\label{fig:fig_3}
\end{figure*}

\subsection{Chern number}\label{Chern number}
To substantiate the phase diagram determined by the correlation lengths, we further calculate the Chern number in different phases by the iDMRG method using a charge pumping scheme~\cite{Laughlin1981}. Specifically, one can envision that if a magnetic flux $\theta$ is adiabatically inserted along the $x$-axis, charges would be pumped from the left to the right side of a cylinder. The accumulative discrepancy of the charges between two sides of the cylinder can be obtained with
\begin{eqnarray}
Q(\theta) = \sum_p \Lambda^{2}_p (\theta) [Q^{L}_p (\theta) - Q^{R}_p (\theta)],
\end{eqnarray}
where $\Lambda_p$ is the singular value obtained after decomposing the whole cylinder into two semi-infinite parts, $Q^{L/R}_p$ is the charge degree of freedom on the left/right side marked for the $p$-th renormalized basis, and $p$ runs over all truncated bond dimension $m$.

Upon inserting the magnetic flux from $\theta$ to $\theta + \delta \theta$, all eigenstates of the system undergo adiabatic evolution, meaning that the system stays at its targeted state, indicated by the fidelity $F(\theta) = \left\lvert \braket{\Psi_0(\theta) \vert \Psi_0(\theta+\delta\theta)} \right\rvert$ being approximately fixed at $1$. As a result, the Chern number can be calculated using the accumulated change with
\begin{eqnarray}
C = \frac{1}{2} \left\lvert Q(2\pi) - Q(0) \right\rvert\ .
\end{eqnarray}

\subsection{Phase transitions}\label{Phase transitions}
We begin by presenting the phase diagram of HHM in the space of parameters $(U, \delta)$, as shown in Fig.~\ref{fig:fig_3}(a). The phase boundaries are determined using the correlation lengths and the Chern number. In particular, we note that the phase boundaries for cylinders ZT$3$-$1$ and tZT$4$-$2$ are fairly consistent. Four phases exist: the CI$_1$ phase with a Chern number $C=1$ and the CI$_2$ phase with a Chern number $C=2$, when $U/2$ and $\delta$ are comparable; a CDW and an SDW insulator phase, both with a zero Chern number, when either $\delta$ or $U/2$ prevails. Other methods have also predicted such phases, e.g., MF, ED, DMFT~\cite{Vanhala2016}, and BDMC~\cite{Tupitsyn2019}, although quantitative agreement regarding the locations of the phase transitions was lacking.

Several remarks on Fig.~\ref{fig:fig_3}(a) are in order: i) The CI$_2$ phase can extend to the lobe at $U/t\approx 11$ and $\delta/t\approx 4.6$, in closer quantitative agreement to the DMFT outcomes~\cite{Vanhala2016}. In addition, for $10 \le U/t \le11$, the CI$_2$ phase survives in the middle of CI$_1$, exhibiting a reentrant behavior. ii) The CI$_1$ phase exists even at the small-$U$ region, making the transition between the CI$_2$ and CDW phase indirect, in qualitative agreement with the ED results~\cite{Vanhala2016}. iii) At $\delta=0$, the transition between the CI$_2$ and SDW phase occurs around $U/t\sim6.8$, comparable to those of ED, DMFT, and BDMC~\cite{Vanhala2016, Tupitsyn2019}. However, when increasing $U$ with a small nonzero $\delta$, both ED and BDMC predict two consecutive transitions from CI$_2$ to CI$_1$, and then to SDW, instead of the CI$_2$-SDW transition predicted by us.

Having established those main features unifying the picture for the phase diagram obtained from different methods, we now illustrate how the phase boundaries in Fig.~\ref{fig:fig_3}(a) are determined by the correlation lengths introduced in Sec.~\ref{Correlation lengths}. In Fig.~\ref{fig:fig_3}(b), the charge correlation length $\xi_{c}$ exhibits two pronounced peaks for $U/t=18$ when sweeping the staggered potential $\delta$, indicative of the critical points of the phase transitions SDW-CI$_1$ and CI$_1$-CDW. In Fig.~\ref{fig:fig_3}(c) for $U/t=8$, one more feature emerges compared to those in Fig.~\ref{fig:fig_3}(c): $\xi_n$ and $\xi_s$ overlap for $\delta/t \lesssim 1.68$ and they exhibit a sharp discontinuity at this point. This is a signature of the phase transition SDW-CI$_2$, since in the SDW phase, the spin and neutral excitation form a spin-triplet excitation.

By computing such correlation lengths for different sets of parameters, we compile the final phase diagram, Fig.~\ref{fig:fig_3}(a). Lastly, we note that despite minor quantitative differences, the results for the smaller cylinder geometry ZT$3$-$1$ [dashed lines in Figs.~\ref{fig:fig_3}(a), (b), and (c)] result in similar behavior to the ones for the tZT$4$-$2$ (solid lines).

After computing the correlation lengths and obtaining the phase diagram, our analysis progresses towards determining the Chern number in different phase regions with the charge pumping scheme introduced in Sec.~\ref{Chern number}; results for the tZT$4$-$2$ cylinder geometry are summarized in Fig.~\ref{fig:fig_3}(d). Specifically, in the case of $U/t=8$ and $\delta/t=1.4$, the adiabatic insertion of a $2\pi$-flux pumps a charge $\Delta Q = Q(2\pi)- Q(0)=0$, confirming that the Chern number is given by $C=0$, which is indicative of the system being in a topologically trivial SDW phase. On the other hand, for $U/t=8$ and $\delta/t=2.9$, a continuous curve with an adiabatic process yields a pumping charge $\Delta Q =4$, which corresponds to a topologically nontrivial phase with a Chern number of $C=2$. Lastly, the CI$_1$ phase is exemplified in the set of parameters $U/t = 15$ and $\delta/t=7.1$ -- a clear accumulated charge $\Delta Q =2$ exposes the CI$_1$ ground-state.

Overall, the phase characteristics revealed by the Chern number calculations for different parameters agree with the expected correlation lengths. Therefore, the phase diagram of HHM can be accurately identified through the combined analysis of correlation lengths and Chern numbers with the iDMRG method.

\subsection{Correlation-length spectrum}\label{correlationlengthspectrum}
Further characterization can be performed by describing the low-energy excitations with momentum resolution, achieved via a particular type of adiabatic driving. For the charge excitations, for example, we add a uniform phase factor in front of hopping terms $\hat{c}^\dag_{l\sigma} \hat{c}^{\phantom{\dag}}_{l'\sigma}$ for both species, which equivalently threads a gauge flux $\theta$ through the thin and infinitely-long cylinder adiabatically. For the cylinder tZT$L_y$-$1$, the corresponding momentum cutting lines, mentioned in Eq.~\eqref{eq:momentum_selection} of Sec.~\ref{MODEL AND METHOD}, follow a set of equations
\begin{eqnarray}
k^{\cal S}_{j,1} = k^{\cal S}_j + \theta / L_y \ \text{and} \ k^{\cal S}_{j, 2} = L_y k^{\cal S}_j \ \text{modulo} \ 2\pi\ , \quad \label{charge excitations}
\end{eqnarray}
which sweeps the FBZ as $\theta$ grows from $0$ to $2\pi$. Here, $k^{\cal S}_{j,1}$ and $k^{\cal S}_{j,2}$ denote the momenta in the basis of two reciprocal vectors $\bm{b}_1$ and $\bm{b}_2$. As a result, we can obtain the momentum-dependence features of the correlation-length spectrum $1 / \xi_c$ for the charge excitations in the subspace $(1,1/2)$ of the associated transfer matrix.

\begin{figure}[t]
\centering
\includegraphics[width=\linewidth]{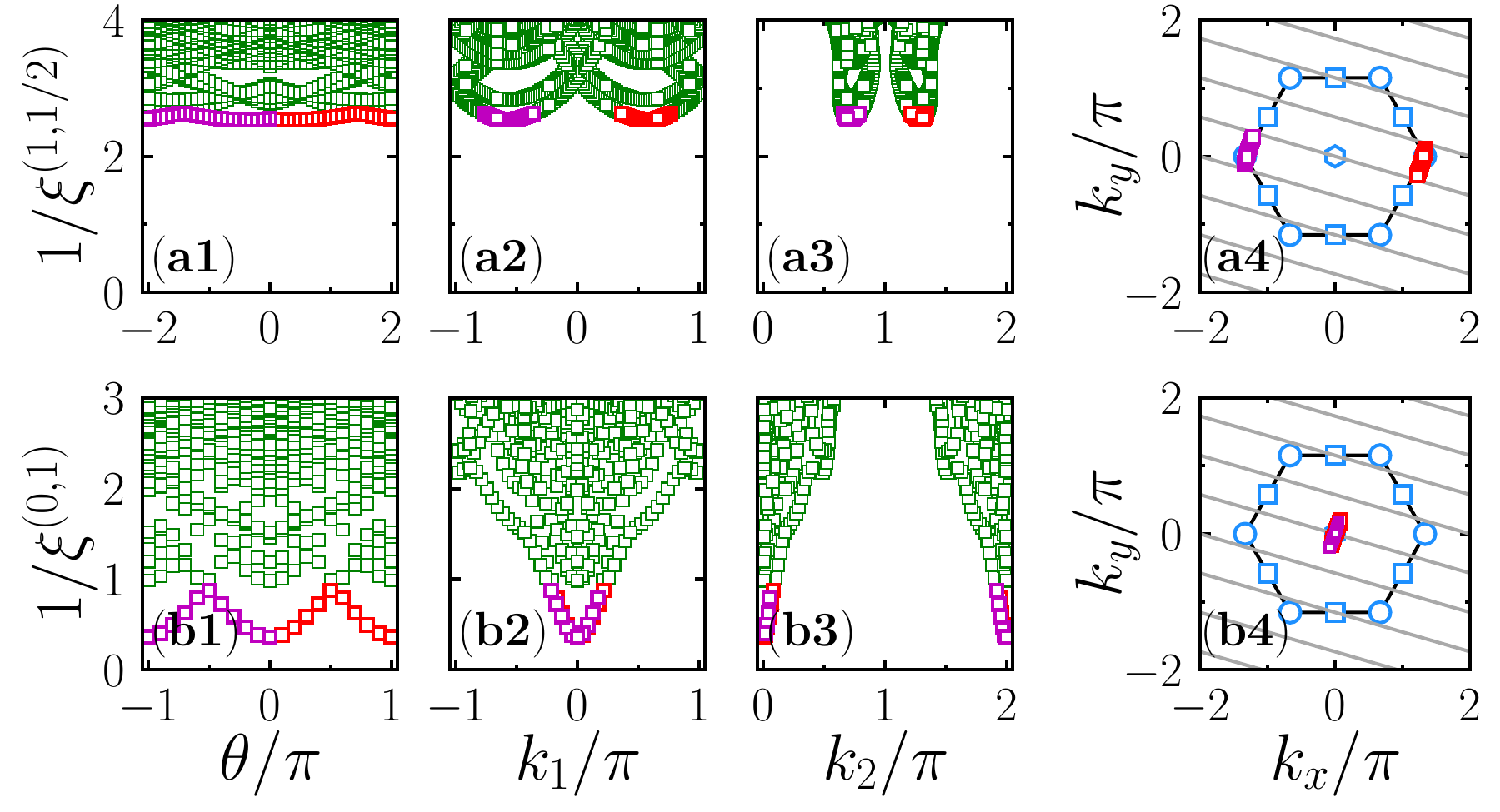}
\caption{The correlation-length spectrum as a function of the inserted flux $\theta$, the momenta $k_1$ and $k_2$ in the reciprocal vectors $\bm{b}_1$ and $\bm{b}_2$, respectively. We consider two distinct excitations for the cylinder tZT$4$-$1$ when $U / t = 15$ and $\delta / t = 0$ in the SDW region. (a.1-3) The inverse correlation lengths $1 / \xi_c$ for the charge excitations in the subspace $(1,1/2)$ of the transfer matrix. The lowest-energy charge excitations in the positive (red) and negative (magenta) branches hit $K$- points $\mathbf{K}^\pm$ as $\theta$ approaches $\pm 2\pi/3$, individually. (b.1-3) The inverse correlation lengths $1 / \xi_s$ for the spin-flipping excitations in the subspace $(0,1)$. Both of the lowest-energy spin excitations hit the $\Gamma$-point as $\theta$ approaches $0$ and $\pm \pi$. (a.4, b.4) We annotate the momenta of the lowest-energy excitations for various $\theta$, color marked according to the previous panels. The momentum-cutting lines (grey) for the charge and spin-flipping excitations are indicated by Eq.~\eqref{charge excitations} and Eq.~\eqref{spin excitations}, respectively, when $\theta = 0$. Here, we set the truncated bond dimension $m=4096$.
}\label{fig:fig_7}
\end{figure}

In contrast, to produce the spectrum for the spin-flipping excitations in the subspace $(0, 1)$, it is necessary to adiabatically thread a $-\theta$ ($+\theta$) flux to the cylinder for the spin-up (spin-down) electrons. Considering the exchanging symmetry between two spin species, it is sufficient to increase $\theta$ from $0$ to $\pi$. Similarly, for the cylinder tZT$L_y$-$1$,  the corresponding momentum is resolved as
\begin{eqnarray}
k^{\cal S}_{j,1} = k^{\cal S}_j - 2 \theta / L_y \ \text{and} \ k^{\cal S}_{j,2} = L_y k^{\cal S}_j \ \text{modulo} \ 2\pi\ . \quad \label{spin excitations}
\end{eqnarray}

\begin{figure}[!t]
\centering
\includegraphics[width=\linewidth]{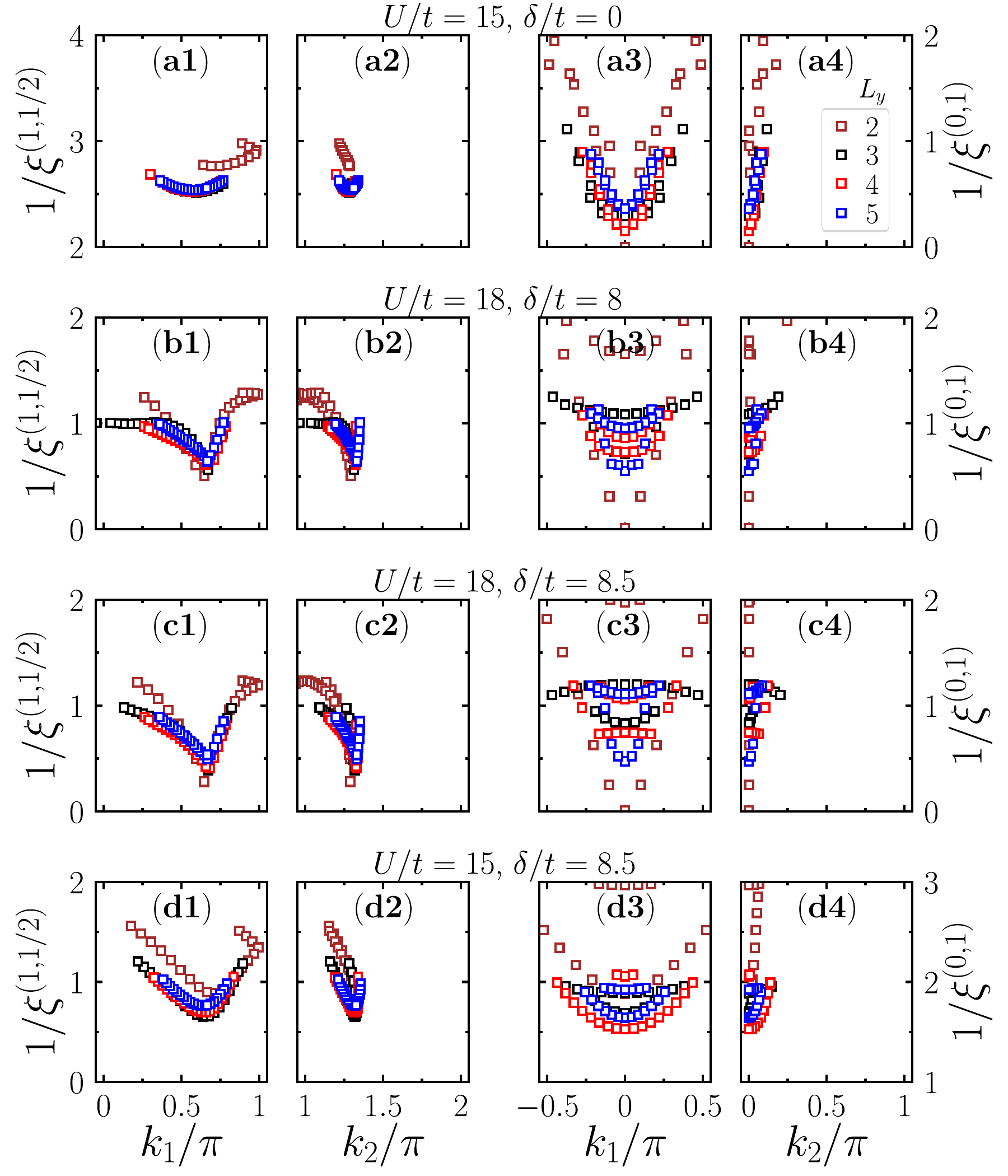}\\
\caption{Finite circumference effect of the lowest-lying correlation-length spectrum level $\xi^{(1,1/2)}$ for the charge excitation (two left columns), and $\xi^{(0,1)}$ for the spin excitation (two right columns). We choose four typical parameter sets: (a1-4) $U=15$ and $\delta=0$, (b1-4) $U=18$ and $\delta=8$, (c1-4) $U=18$ and $\delta=8.5$, and (d1-4) $U=15$ and $\delta=8.5$. Besides, we systematically select four circumferences: $L_y=2$ (brown), $3$ (black), $4$ (red), $5$ (blue) in the cylinder tZT$L_y$-$1$. The truncated bond dimension is given by $m=4096$.}
\label{fig:fig_8}
\end{figure}

First, we exemplify the correlation-length spectrum for a typical parameter $U / t =15$ and $\delta / t = 0$ in the SDW phase region. As shown in Fig.~\ref{fig:fig_7}(a1), the lowest-lying branch of charge excitations reaches its minimum at $\theta=\pm 2\pi/3$, where one of the momentum cutting lines goes through either of
\begin{eqnarray}
\mathbf{K}^\pm = \left( k_1 = \pm \frac{2\pi}{3},\ k_2 = \pm \frac{4\pi}{3} \right) \ \text{modulo} \ 2\pi\ ,
\end{eqnarray}
which equivalently corresponds to the momenta $k_x=\pm 4\pi/3$ and $k_y=0$ in Cartesian coordinates, marked in Fig.~\ref{fig:fig_7}. For the spin-flipping excitations, the lowest-lying branch, as shown in Figs.~\ref{fig:fig_7}(b1-3), gets the minimum at the high-symmetry point $\mathbf{\Gamma} = (0, 0)$ modulo $2\pi$ as $\theta=0$. The cone structure in the vicinity of the $\Gamma$-point implies the emergence of Goldstone modes, providing a hallmark of gapless excitations in the SU($2$)-symmetry broken N\'{e}el phase~\cite{Bernu_1992, Bernu_1994}. The small but finite $L_y / \xi_s$ at the $\Gamma$-point is caused by the finite truncated bond dimension $m=4096$ and circumference $L_y$.

Second, we check the effects of the finite circumference of the cylinder on the spectrum as we vary $L_y$ from 2 to 5, while choosing a sufficiently large $m \le 4096$. In the SDW phase region, at $U / t = 15$ and $\delta / t = 0$, as shown in Figs.~\ref{fig:fig_8}(a1-2), all minima of the lowest-lying branches of charge excitations are located at the $K$-points $\mathbf{K}^\pm$ and $1/\xi_c$ approaches a constant as $L_y$ grows. In turn, for the minima of spin-flipping excitations at $\mathbf{\Gamma}$, shown in Figs.~\ref{fig:fig_8}(a3-4), the inverse correlation length $1/\xi_s$, theoretically associated to the gapless Goldstone modes, is expected to decrease as the circumference $L_y$ grows. We notice that the approach to the thermodynamic limit is highly influenced by large finite-size effects when $L_y \le 5$. In contrast, in the CDW phase, at $U / t = 15$ and $\delta / t = 8.5$, all the minima of the inverse correlation lengths $1 / \xi_c$ and $1 / \xi_s$ for the charge and spin-flipping excitations in Figs.~\ref{fig:fig_8}(d1-4) approach convergence as the circumference $L_y$ increases.

Lastly, in the CI$_1$ region, such as $\delta / t = 8$ and $8.5$ when $U / t = 18$, the equivalent charge gap $\Delta_c \propto 1/\xi_c$ keeps finite [Figs.~\ref{fig:fig_8}(b1-2) and \ref{fig:fig_8}(c1-2)]. Up to relatively small $L_y=5$, we can not conclude that the gap for the spin-flipping excitations closes in the thermodynamic limit of infinite cylinder circumference [Figs.~\ref{fig:fig_8}(b3-4) and \ref{fig:fig_8}(c3-4)]. Yet, compared to the spin excitation spectrum in the SDW phase for $U / t = 15$ and $\delta / t = 0$ [Figs.~\ref{fig:fig_8}(a3-4)], we find that the low-energy spectrum of the spin-flipping excitations in the CI$_1$ phase region has a similar qualitative profile, based on which we believe that the resulting ground state behaves as an N\'{e}el state with SU($2$) symmetry breaking. However, when the iDMRG simulations are applied to a cylinder with a finite circumference $L_y$, we consistently obtain the lowest-energy state with a total spin of $0$. This is because the energy spacing between quasi-degenerate joint states in distinct total spin sectors is approximately proportional to $1 / L_y$~\cite{Bernu_1992, Bernu_1994}. The resulting spin-spin correlation functions preserve the spin rotation symmetry, as shown in Fig.~\ref{fig:appendix_ssc} in Appendix~\ref{app:sosym_app} for the cylinder circumferences studied.

\section{Summary and conclusions}\label{CONCLUSIONS}

We have investigated the ground-state phase diagram of HHM at half-filling using the MF and iDMRG methods. Within MF, we examine four physical quantities: the effective mass term for each spin species, the staggered magnetization, the sublattice density imbalance, and the excitation gap, extracting a phase diagram consistent with previous MF calculations~\cite{Vanhala2016}. Remarkably, the calculated effective mass term, as a function of the interaction $U$, clearly demonstrates a scenario where only one of the two spin species contributes to the Chern number, thereby unequivocally identifying the CI$_1$ phase.

Within an unbiased scheme using iDMRG, we have employed the Chern number and three different characteristic correlation lengths to determine the phase diagram in the $(U,\delta)$ space, with $\pm\delta$ the staggered potential energies of the two sublattices of the honeycomb lattice. In addition, we unveil a detailed analysis of the low-energy charge and spin excitations by resolving the momentum information in the correlation-length spectrum. Our state-of-the-art phase diagram not only analyzes the discrepancies between the phase boundaries predicted by different approaches as MF, ED, DMFT~\cite{Vanhala2016}, and BDMC~\cite{Tupitsyn2019}, but also excludes the emergence of any new phase other than the SDW, CDW, CI$_1$, and CI$_2$ phases, therefore establishing a reliable benchmark for future theoretical and experimental investigations of HHM.

\begin{acknowledgments}
We thank Wei Pan, Ji-Lu He, and Tian-Cheng Yi for the helpful discussions. We acknowledge funding from the Ministry of Science and Technology of the People's Republic of China (Grant No.~2022YFA1402700) and the National Natural Science Foundation of China (Grants No.~U2230402). R.~M.~acknowledges support from the NSFC Grants No.~12111530010, 12222401, and No.~11974039. H.-G.~L. acknowledges funding from the NSFC Grants No.~11834005 and No.~12247101. X.~Q.~W.~acknowledges funding from the NSFC Grant No.~11974244. S.~H. acknowledges funding from NSFC Grant No.~12174020. The computations were performed on the Tianhe-2JK at the Beijing Computational Science Research Center (CSRC) and the Quantum Many-body {\rm I} cluster at the School of Physics and Astronomy, Shanghai Jiaotong University.
\end{acknowledgments}

\appendix
\label{APPENDIX}

\renewcommand\thefigure{S\arabic{figure}}

\setcounter{figure}{0}

\section{tZT2-1 cylinder}\label{app:TZT2-1}

In Sec.~\ref{iDMRG calculations}, we have presented a detailed account of the quantitatively consistent phase diagrams of ZT$3$-$1$ and tZT$4$-$2$ with the iDMRG method. In this section, we extend our investigation to the cylinder tZT$2$-$1$ and demonstrate the significance of the proper choice of the lattice structure in the iDMRG simulation.
\begin{figure}[!htbp]
\centering
\includegraphics[width=0.48\textwidth]{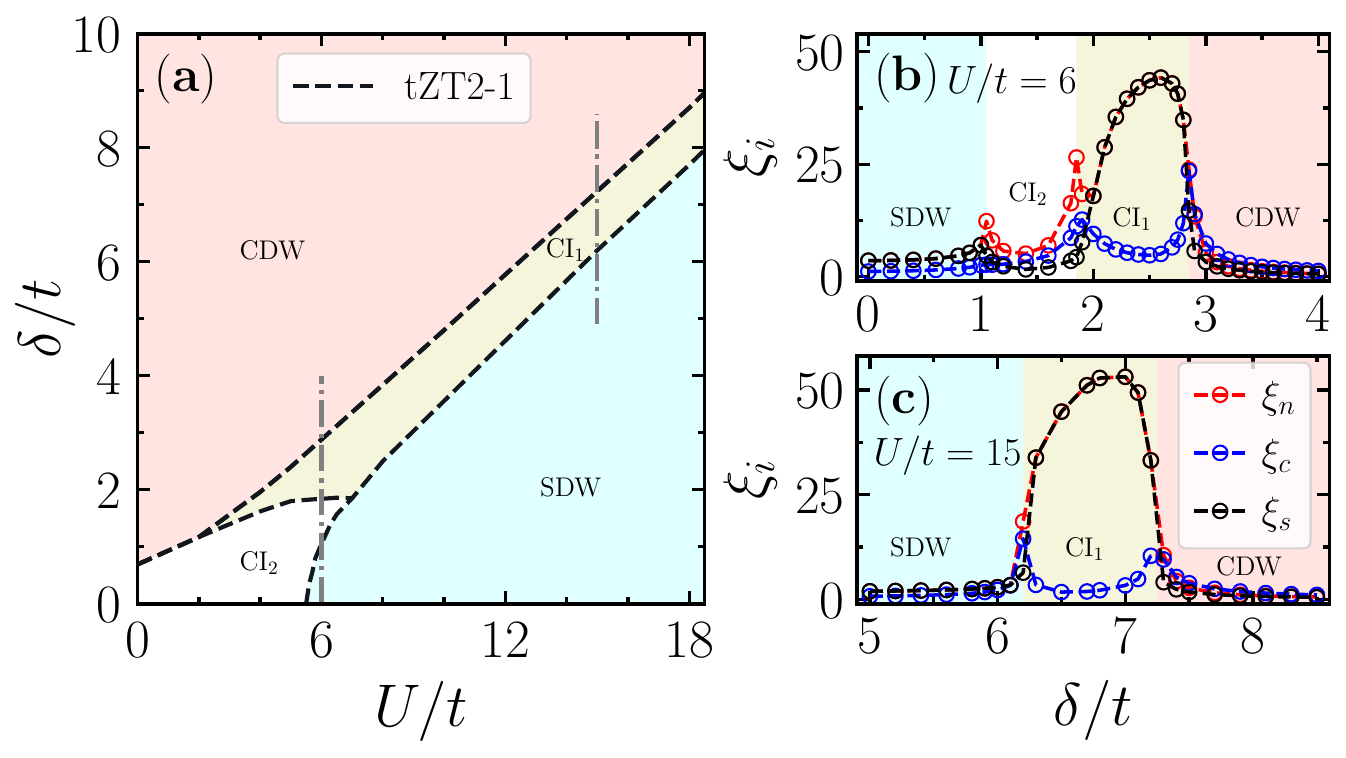}\\
\caption{Phase diagram of the HHM in the space $(U,\delta)$ of parameters for cylinder tZT$2$-$1$. Unlike in Fig.~\ref{fig:fig_3}(a), the CI$_2$ phase does not exhibit a reentrant behavior in the CI$_1$ phase, a feature to which we attribute being related to finite-size effects -- see text. (b,c) Correlation lengths as a function of the potential energy $\delta$ by setting onsite interaction (b) $U/t=6$ and (c) $15$. All results are obtained by the iDMRG simulations with a truncated bond dimension $m=2048$.}
\label{fig:fig_4}
\end{figure}

To summarize these results, Fig.~\ref{fig:fig_4}(a) shows the phase diagram in the $(U, \delta)$ space for the tZT$2$-$1$ cylinder. The phase diagram is qualitatively similar to the one before, exhibiting four different phases. The most remarkable difference between Fig.~\ref{fig:fig_3} and Fig.~\ref{fig:fig_4}(a) is the direct transition between CI$_2$ and SDW, which leads to the absence of CI$_2$ phase when $U/t=10\thicksim11$. In other words, the CI$_2$ phase misses the reentrant behavior in the CI$_1$ phase. Ref.~\cite{Yuan2023} obtained the precise same result on the $12$A cluster. Again, we attribute this difference to a finite-size effect, i.e., $L_y=2$ is too small to adequately capture fine features of the phase diagram in the thermodynamic limit. Overall, an accurate selection of the lattice structure on the infinite cylinder in our work effectively preserves the physical properties of the ground state related to the high-symmetry points.

Figure~\ref{fig:fig_4}(b) and (c) shows various correlation lengths at $U/t=6$ and $15$, respectively, which qualitatively agrees with the conclusions in the large-$U$ region for the cylinder ZT$3$-$1$ mentioned in the main text. We also obtained the SDW-CI$_1$ and CI$_1$-CDW transition points by observing the peaks of the correlation length for the lowest-energy charge excitation $\xi_c$. The correlation lengths for the spin and neural excitations exhibit an ``extra degeneracy'' of $\xi_s = \xi_c$ in the CI$_1$ state. This implies the presence of an excitation ``magnon" with degenerate $S^z = 0$, $\pm1$ states in the lowest excited energy level. On the other hand, the multiplet vanishes in the CI$_2$ state. Hence, we distinguish the CI$_1$ and CI$_2$ states by examining the overlaps between $\xi_s$ and $\xi_n$.

\begin{figure}[!t]
\centering
\vskip0.1in
\includegraphics[width=0.4\textwidth]{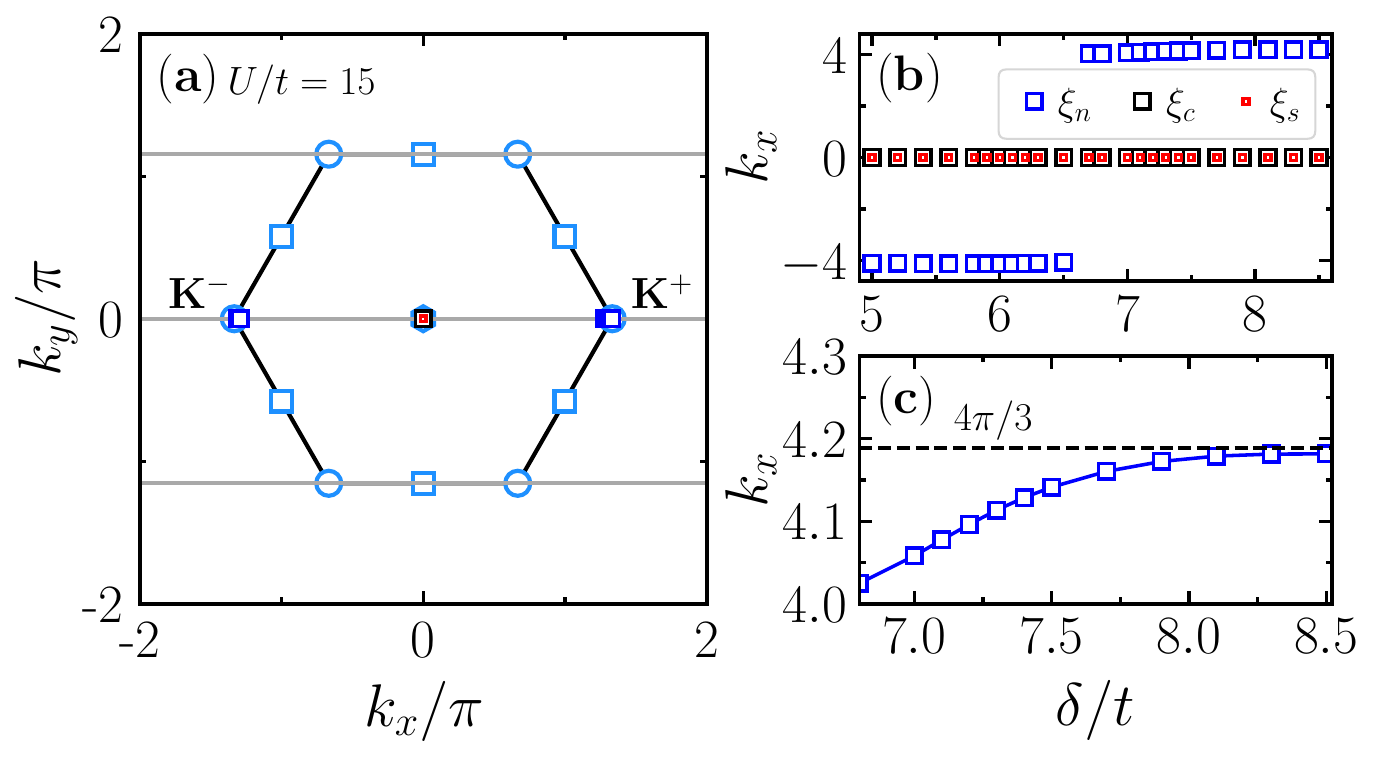}\\
\caption{(a) Excitation momentum in the first Brillouin zone -- several excitation momenta are superimposed and color-coded according to (b). (b) The corresponding excitation momentum $k_x$ in the x-axis as a function of the potential energy $\delta$ by setting onsite interaction $U/t=15$ for cylinder tZT$2$-$1$. (d) A zoom-in of the excitation momentum $k_x$ after it transitioned to close $\mathbf{K}^+$, the x-axis projection of which is equal to $4\pi/3$ (black dashed line). We use a truncated bond dimension $m=2048$ in the iDMRG simulations.}
\label{fig:appendix_idmrg}
\end{figure}

For the lattice structure under consideration, the allowed momentum values appear as a discrete set of lines tangent to two edges of FBZ [see Fig.~\ref{fig:appendix_idmrg}(a)], owing to the finite number of sites on the circumference of the cylinder. Notably, despite being a lattice encompassing the inequivalent Brillouin zone corners, it does not capture other essential symmetry points, as it misses some of the inequivalent $M$ points.

The x-axis projection of the excitation momentum $k_x$ in FBZ with increasing $\delta$ is illustrated by the empty blue hexagons positioned near the momentum points $\mathbf{K}^-$ and $\mathbf{K}^+$. The momentum associated with the spin and neutral excitations, represented by red and black empty hexagons, respectively, are located at the $\mathbf{\Gamma}$-point instead. Generally, the momentum $k_x$ of the low-energy excitations along the longitudinal direction of the cylinder can be obtained according to Eq.~\eqref{tmeigval}.

To give an intuitive understanding of the low-energy excitation momentum points, Figs.~\ref{fig:appendix_idmrg}(b) and (c) show the $k_{x}$ dependence of $\delta$. The momentum point jumps from $\mathbf{K}^-$ to $\mathbf{K}^+$ within the CI$_1$ phase region in Fig.~\ref{fig:appendix_idmrg}(b). We have also checked the phase transitions SDW-CI$_1$/CI$_2$ related to the $\mathbf{K}^+$ point, while the others are about $\mathbf{K}^-$, by sweeping the parameter space. Figure~\ref{fig:appendix_idmrg}(c) gives a zoom-in of the excitation momentum for the charge excitation around $\mathbf{K}^+$ for $\delta/t\gtrsim 6.8$, close to where the momentum point jumping occurs. Unlike the MF results in the inset of Fig.~\ref{fig:fig_2}(c), the energy gap closure here does not strictly happen at the Dirac point. We emphasize that, similarly, for ED studies in small clusters, charge excitation momentum may not coincide with the $K$-points~\cite{Yuan2023}. As here, such $12$-site clusters (often labeled as $12$A) do not present all point-group symmetries also missing many of the inequivalent $M$-points.

\begin{figure}[!t]
\centering
\includegraphics[width=0.96\columnwidth]{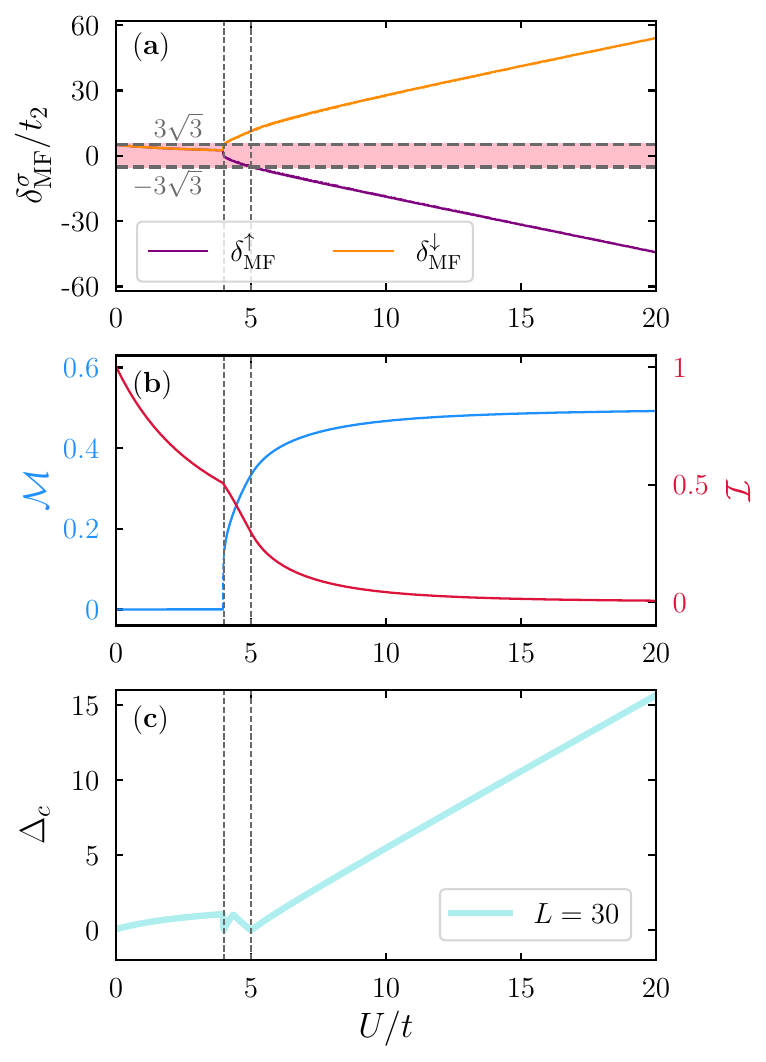}\\
\caption{Mean-field results for $\delta/t = 1$, in a lattice with $L=30$. (a) The effective staggered potential for each spin species after the interactions are accounted for on a mean-field level; the shaded region marks the topological regime. (b) the staggered magnetic order parameter, the sublattice density imbalance, and the excitation gap (c) vs.~$U/t$.}
\label{fig:appendix_mf}
\end{figure}

\begin{figure}[!t]
\centering
\includegraphics[width=0.46\textwidth]{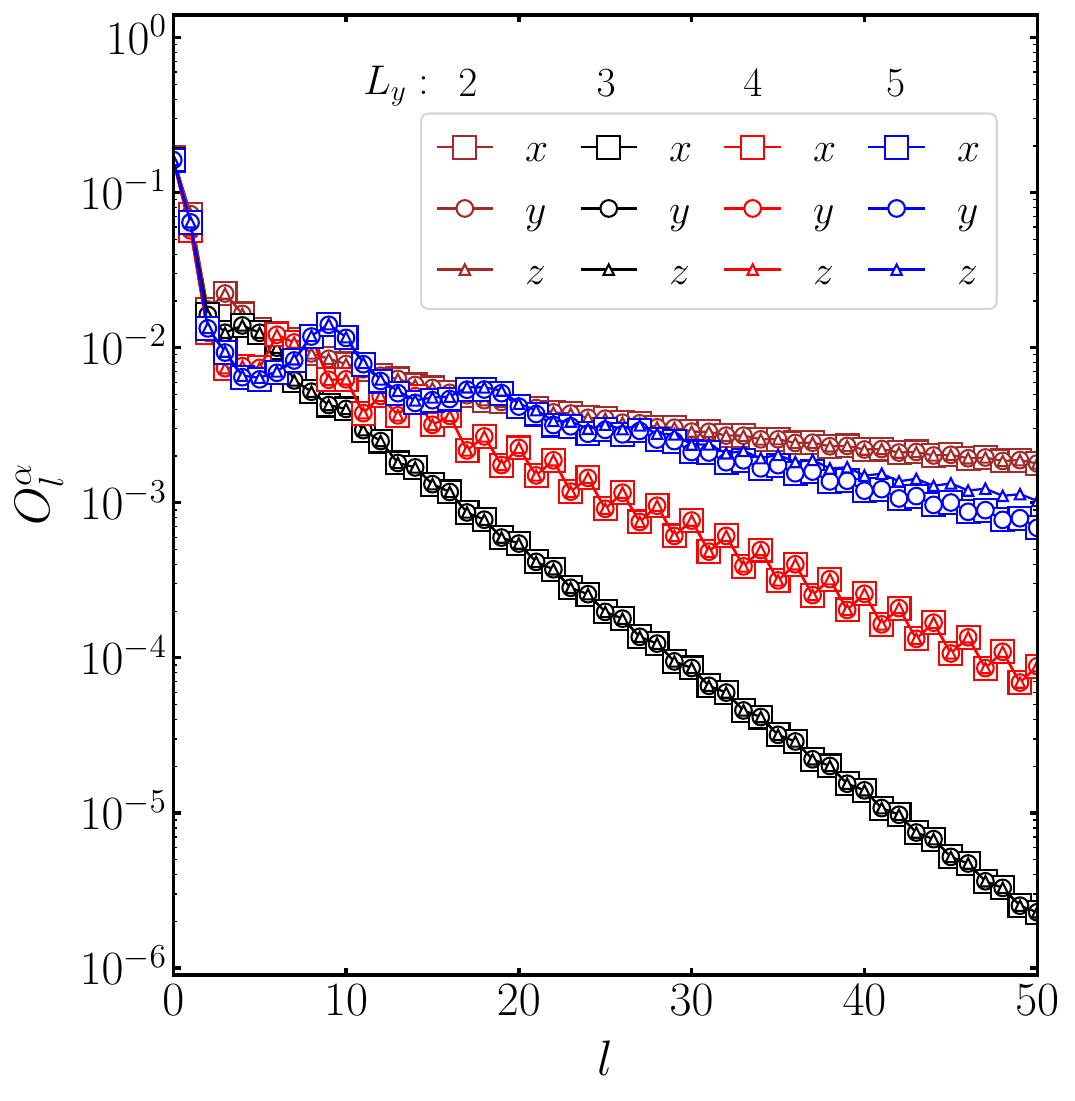}\\
\caption{Spin-spin correlation function $O^\alpha_l = \braket{\hat{S}^\alpha_0 \hat{S}^\alpha_l}$ ($\alpha=x$, $y$, and $z$) for the cylinder tZT$L_y$-$1$ with different circumferences, i.e., $L_y=2$ (brown), $3$ (black), $4$ (red), $5$ (blue). We choose $U/t=18$ and $\delta/t=8$ in the CI$_1$ phase region, which share the same parameter set of Fig.~\ref{fig:fig_8}(b1-4). All results are obtained by the iDMRG simulations with a truncated bond dimension $m=4096$.}
\label{fig:appendix_ssc}
\end{figure}

\section{Other $\delta$ values in the mean-field theory} \label{app:mf_app}

The main text shows a line cut in the $(U, \delta)$ phase diagram using the MF method, with $\delta/t = 6$ fixed. While illustrative of the CI$_1$ phase and the topologically trivial ones, either band-insulating CDW or SDW, it does not show the fourth phase in the phase diagram, corresponding to CI$_2$. For completeness, we include in Fig.~\ref{fig:appendix_mf} data for $\delta/t=1$, which with increasing interactions $U$, drive the system across the phases CI$_2$, CI$_1$, and SDW. As for the $\delta/t=6$ case shown in the main text, the topological character of the different phases can be readily inferred from the value of the effective staggered potential $\delta_\text{MF}^\sigma$ [Fig.~\ref{fig:appendix_mf}(a)], checking whether $|\delta_\text{MF}^\sigma/t_2|<3\sqrt{3}$. For small-$U$, both $\delta_\text{MF}^\uparrow$ and $\delta_\text{MF}^\downarrow$ satisfy this condition, and the resulting phase exhibits a Chern number of $2$. The spontaneous symmetry-breaking that occurs when $\delta_\text{MF}^\uparrow \neq \delta_\text{MF}^\downarrow$ is marked as a first-order transition with a discontinuous magnetic order parameter $M$ [Fig.~\ref{fig:appendix_mf}(b)], whereas a continuous transition occurs from the CI$_1$ to the SDW regime. Nevertheless, in both cases, the gap excitation $\Delta_c({\bf k}) \to 0$ identifies the topological transition.

\section{Spin-rotation symmetry} \label{app:sosym_app}

Here, we present the spin-spin correlation function $O^\alpha_l = \braket{\hat{S}^\alpha_0 \hat{S}^\alpha_l}$ ($\alpha=x$, $y$, and $z$) along a ``snake-like" path for the cylinder tZT$L_y$-$1$ in the CI$_1$ phase region. We use a parameter set of $U/t=18$ and $\delta/t=8$. For four different circumferences $L_y=2$, $3$, $4$, and $5$, we demonstrate that the gapless lowest-energy spin excitations, indicated by the inverse of the correlation lengths in Fig.~\ref{fig:fig_8}(b3-4), do not arise from spin-rotation symmetry breaking at these lattice sizes. Also note that the decay rate, non-monotonic in $L_y$, is tied to the non-monotonic spin gap structure observed in Fig.~\ref{fig:fig_8}(b3-4).

\bibliography{ref}

\end{document}